\documentclass{article} 
\usepackage{graphicx}        % standard LaTeX graphics tool
\begin{document}
\centerline{\bf Love kills: Simulations in Penna Ageing Model}
\medskip

\noindent
Dietrich Stauffer$^1$, Stanis{\l}aw Cebrat, Thadeu J.P. Penna$^2$ and 
A. O. Sousa$^3$
\medskip

\noindent
Department of Genomics, 
Wroc{\l}aw University, ul. Przybyszewskiego 63/77, 51-148 Wroc{\l}aw, Poland

and 

\noindent
Laboratoire PMMH, \'Ecole Sup\'erieure de Physique et de Chimie
Industrielles, 10 rue Vauquelin, F-75231 Paris, France
\medskip

\noindent
$^1$ visiting from Institute for Theoretical Physics, Cologne University,
D-50923 K\"oln, Euroland
\medskip

\noindent
$^2$
Visiting from Instituto de F\'{\i}sica, Universidade
Federal Fluminense; Av. Lito\-r\^{a}nea s/n, Boa Viagem,
Niter\'{o}i 24210-340, RJ, Brazil; and National Institute of Science and 
Technology for Complex Systems, Brazil
\medskip

\noindent
$^3$
Visiting from Departamento de F\'{\i}sica Te\'orica e Experimental,  
Universidade Federal Rio Grande do Norte, 59072-970 Natal, Rio Grande  
do Norte, Brazil

\bigskip

{\small
The standard Penna ageing model with sexual reproduction is enlarged by
adding additional bit-strings for love: Marriage happens only if the 
male love strings are sufficiently different from the female ones. We 
simulate at what level of required difference the population dies out. 

}
\medskip
Love may have been a female invention long ago, when the brain of genus homo
became large and required more food \cite{love}. Thus the father was needed to 
help feed mother and baby. We try to check here if the restriction in the
number of suitable mates caused by love can drive the whole population
to extinction. First we describe the biology, then our model, and finally
our results. 

For sexual reproduction of diploid organisms, usually two individuals of 
different gender are necessary (hermaphrodites are the exception). In 
theoretical concepts, considering a random assortment of genes, any 
preferences in the selection of partners for reproduction usually are 
ignored. In fact, it is known that such a selection can be a very 
important factor for the evolution of the genetic pool of a species. 
However, the selection of potential mates is only the first step 
influencing the assortment of genes. Generally, the assortment of genes 
can be affected at the pre-zygotic stages, like mating preferences or 
gamete preselection, or at post-zygotic stages, like miscarriage, or any 
other differential selection of individuals before and during their 
reproduction period. Sometimes it is difficult to find out at which stage 
the random assortment is disturbed.

The first step generating non-random distribution of genes is the mating 
preference. There are more and more informations suggesting that there 
are not only the obvious phenotypic traits like size or strength of the 
male (or his bank account) which are preferred by females but that there 
are some other genetic characters which influence the mating decisions or 
even tune the genetic relations between mating pairs. Odour of individuals 
is such a trait determined by its genetic configuration. In the generation 
and recognition of an individual's odour at least two groups of genes are 
involved. One group is the Major Histocompatibility Complex (MHC) and the 
second one is a family of Olfactory Receptor (OR) genes (the largest gene 
family in the mammal genomes).

MHC is a large cluster of genes located on the 6th human chromosome 
composed of more than 200 loci involved in different immunological functions
\cite{weisbuch}. The main function of MHC gene products is presenting the 
foreign antigens on the surface of cells. Products of MHC class I are 
present on all nucleated cells of organisms and their role is to present 
in their context a foreign antigen (i.e. peptide of a virus infecting the 
cell). Such a stigmatised cell is recognized and killed by T lymphocyte. 
Products of genes of MHC class II are present on the special group of 
immunological competent cells - ``antigen presenting cells'' involved in 
stimulation the immunological response against the presented antigen.

There are several characteristic genetic properties of the MHC gene 
cluster: Compared with the rest of the genome, the recombination rate is lower,
the mutation rate may be lower for some loci and higher for others,
and other properties less relevant for the present simulations.

It is assumed that those specific properties of the cluster are connected 
with its functions - individuals which possess more different MHC alleles 
can present more different foreign antigens. This is one reason for being 
highly heterozygous in the MHC region. Nevertheless, to keep the high 
level of heterozygosity in the MHC regions other specific properties of 
this region have developed. The most intriguing is the possibility of 
recognizing the configuration of MHC complex of the potential mating 
partner. It was already in 1976 when Yamazaki et al., found out that mice 
heterozygous in MHC loci are more preferred as mating partners than 
homozygous mice. Next, it was found that the fraction of born homozygous 
mice is lower than expected under assumption of random mating \cite{potts}.
It was an effect of both, the non-random mating and biased miscarriage.

Other experiments indicated that mice and rats can recognize partners 
differing in MHC loci \cite{yama79,brown} and 
sometimes even in one locus (i.e. H-2K locus \cite{yama83}). This 
ability of MHC recognition and non-random mating seems to be a more 
general property of many species. It has been found in fish 
\cite{landry,aeschlimann} that they can choose a partner such 
that the probability of producing heterozygous offspring is higher or that 
the two mating partners differ in MHC loci.

The most spectacular finding was that humans also have the ability of 
recognizing the MHC of the partner. It has been found that women prefer 
the odour of men which differ in their HLA-A, HLA-B and HLA-DR alleles 
\cite{wede95}. Next, experiments performed on humans have shown 
that selection prefers combination of partners which have the least number 
of common alleles which renders the highest heterozygosity of the 
offspring \cite{wede97}. (That is why it would help to put
at least four-alleles loci into MHC bitstrings. In such a case two mating 
partners can have different all alleles which ensures that the mother and her 
foetus are also different).  There are at least two non-excluding 
hypotheses explaining the trend for non-random mating and higher 
heterozygosity in the MHC loci. The first one assumes that heterozygous 
individuals can present more foreign antigens what could be especially 
important during multifactor infections \cite{doherty,hughes,penn,mcclelland}.
The second hypothesis assumes that it is a Red Queen effect \cite{langefors}. 
There is a continuous arms race between parasites and hosts. If a high 
fraction of hosts can present the parasite's antigen to their 
immunological system, the parasite has to change its antigen to broaden 
its effective host range. That induces further diversification of MHC 
alleles.

This mechanism could be also important in avoiding the mating between too 
closely related individuals - the cheapest way for generating a higher 
biodiversity.

The partner's MHC recognition is a kind of pre-zygotic selection but there 
is also an early post-zygotic selection connected with MHC. It has been 
observed that if couples share more common alleles in the MHC region, women 
are more prone for early spontaneous abortion [Hedrick 1998]. Probably 
this is connected with an expression of a specific progesterone-induced 
blocking factor (PIBF) which prevents the immunological attack of mother 
against the foetus. If MHC antigens inherited from father differ 
negligibly from those inherited from mother, the PIBF genes are 
under-expressed which allows the mother to develop effective immunological 
response against her foetus \cite{druckmann} and leads to 
the recurrent spontaneous abortion.

This mechanism could be also important in avoiding to give birth to 
relatively highly homozygous offspring, if avoiding the mating between 
too closely related individuals has failed - the next cheapest way for 
generation a higher biodiversity.

If we assume that there are mechanisms providing the non-random assortment 
of MHC alleles, then these phenomena should affect the assortment of other 
genes, at least those linked to the MHC region. What is also interesting, 
the other cluster of genes, the Olfactory Receptor (OR) genes, is closely 
linked to MHC and it could be a cooperation between MHC and OR genes which 
renders this MHC recognition by smell.

\bigskip
P{\c e}kalski \cite{pek} assumed mate selection to be governed by MHC.
In the same spirit now the sexual Penna ageing model \cite{penna} is modified 
to include mate selection by two additional strings of 16 bits each,
for each individual, unrelated to the two usual bit-strings of length 
$L$ containing the age-relevant genome.
Initially, the additional love 
bit-strings are chosen randomly, different for each individual. This model is 
therefore more complicated than the gamete recognition of Cebrat and 
Stauffer \cite{ceb} based on one bit. It has some similarity with the peacock 
tail or bird song simulations in \cite{peacock}. 

During the at most 20 attempts per iteration of a female to find 
a suitable unmarried
male partner, the new ``love'' bit-strings of the male (A and B) and the female
(C and D), which should be dissimilar, are compared. If the difference (as 
defined below) is smaller than a universal love limit, the male is rejected.
Thus with too stringent requirements for love, the female will often not
find a suitable partner within the allowed 20 attempts, stay single during 
this iteration, thus reduce the total number of new babies, and finally lead 
the population to extinction, as indicated for Germany by present trends. 
(With two love strings A and B,
and gametes a and b for the usual parental genome, love string
A is transmitted if haplotype a was selected for the gamete, and love string B 
if haplotype b was selected.) Except when stated otherwise, no crossover and 
no mutation happens in the love strings. 

The difference $d$ between the female and the male she selects should be large
and is defined as follows: For the male (strings A and B) we determine for each
of the 16 bit positions the sum $m_i, \; i=1,2,\dots,16$ of the two love bits: 
zero, one or two. The analogous sums $f_i$ are calculated for the female. Then
$$ d =   \sum_{i=1}^{16} |m_i-f_i| $$
is the difference used in our simulations and varies between 0 and 32. The
extremum $d=32$ is reached if the two mates are fully complementary to each
other in the love strings, and the two love strings within one individual
agree (no heterozygous positions).

We assume the couple to stay together until death does 
them part \cite{fidel}, with the widow or never married woman 
demanding a $d$ not smaller than the love limit.

We start with the standard Penna model \cite{penna} with usually $L=64$, 
a minimum reproduction age $R = 5L/8$, $T = 3$ active mutations kill, $B =
2$ births are attempted per female and iteration, all mutations are recessive, 
both males and females suffer from one (deleterious irreversible) mutation per 
bitstring and iteration, Verhulst deaths are applied to births only with a
carrying capacity $K$ up to 10 million. After 10,000 iterations, when 
a rough age equilibrium has been established, love is switched on, and the
population may die out (Fig.1). Time, up to $10^7$, is measured by the number 
of iterations after love has been switched on. We look at the average $<d>$ as 
well as the squared width $W2 = <d^2> \; - \; <d>^2$. 

\begin{figure}[hbt]
\begin{center}
\includegraphics[angle=-90,scale=0.33]{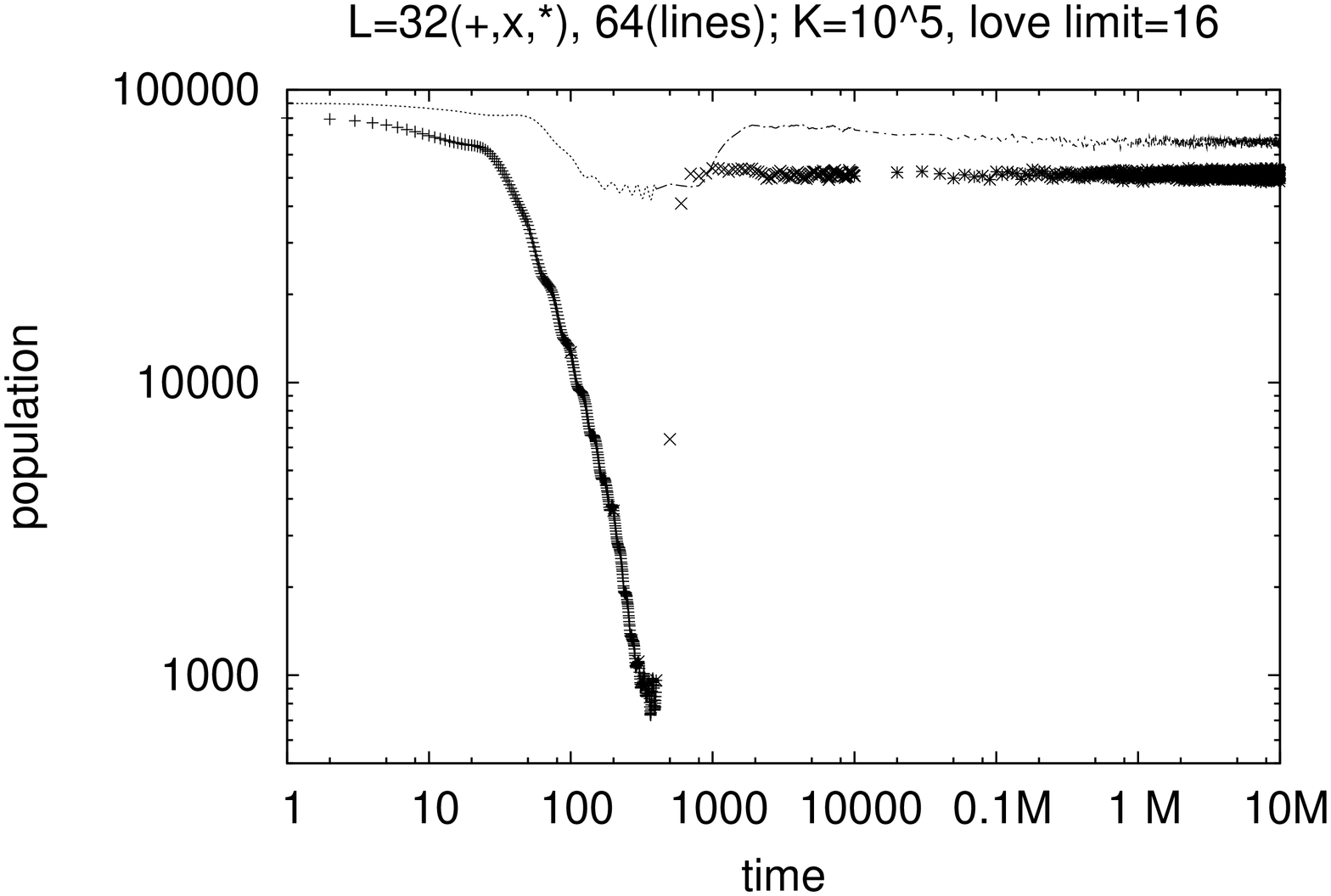}
\includegraphics[angle=-90,scale=0.33]{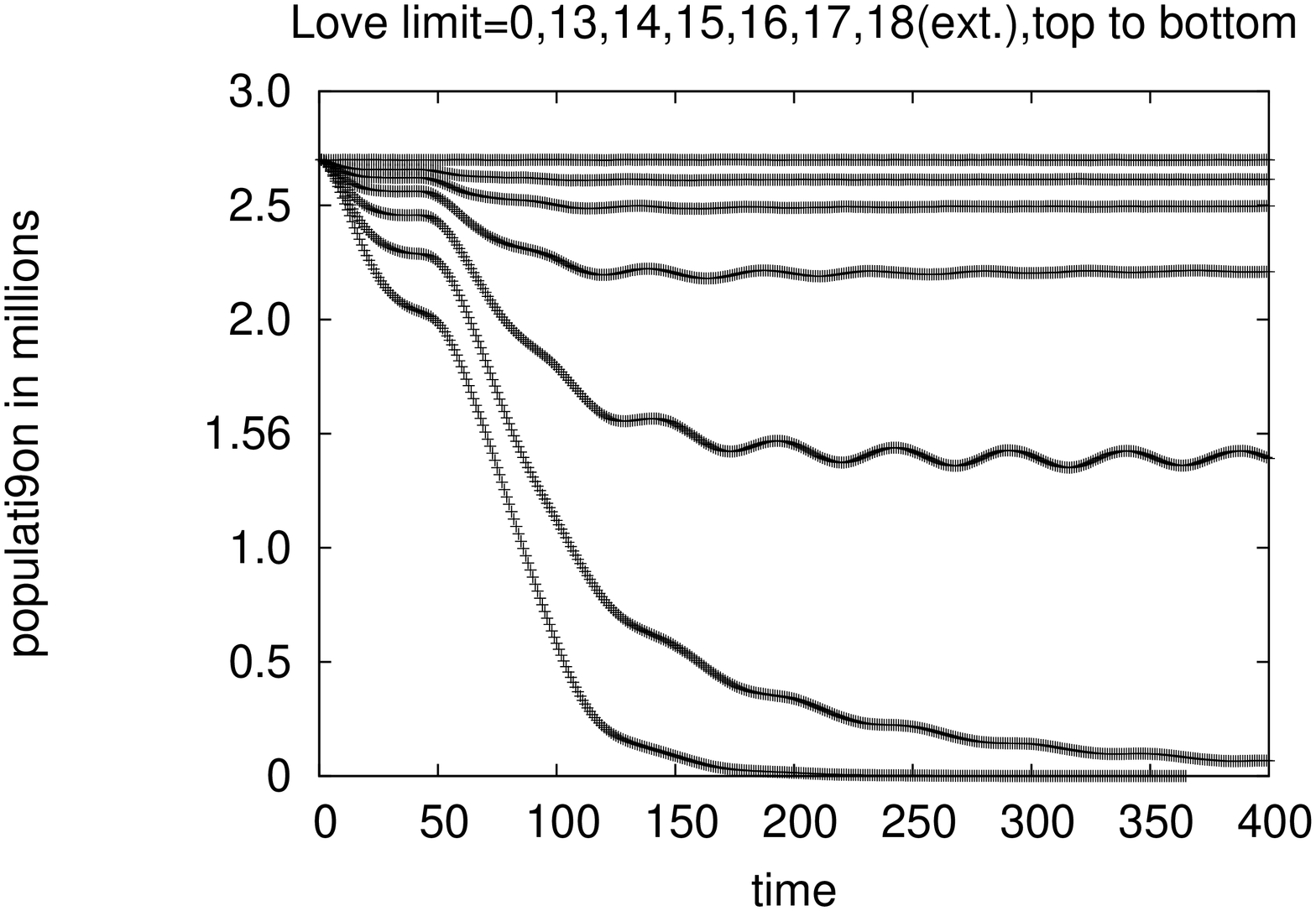}
\includegraphics[angle=-90,scale=0.33]{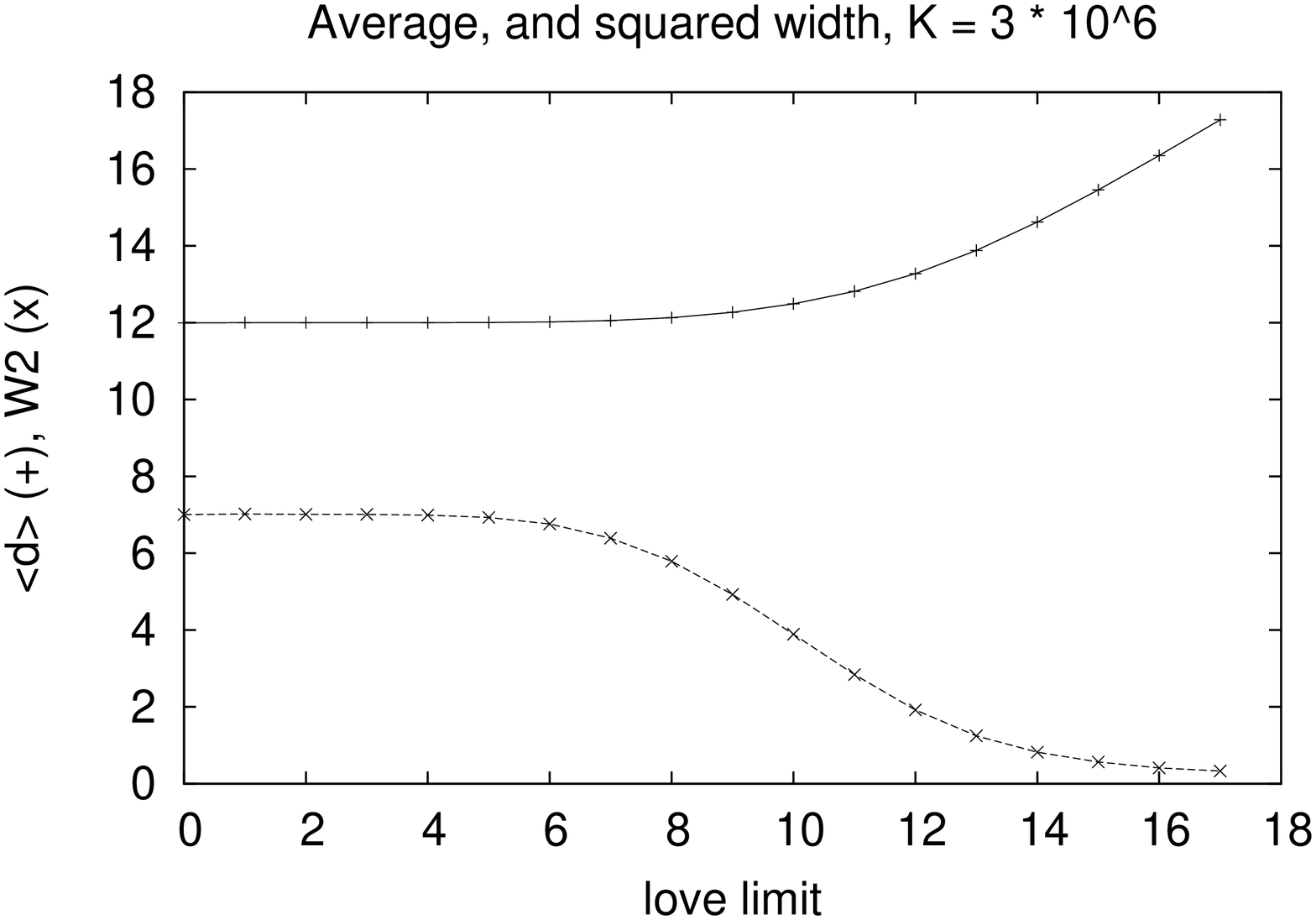}
\end{center}
\caption{
Long-time examples (top), search for extinction limit (centre) and trends with
increasing love limit (bottom). 
}
\end{figure}

\begin{figure}[hbt]
\begin{center}
\includegraphics[angle=-90,scale=0.4]{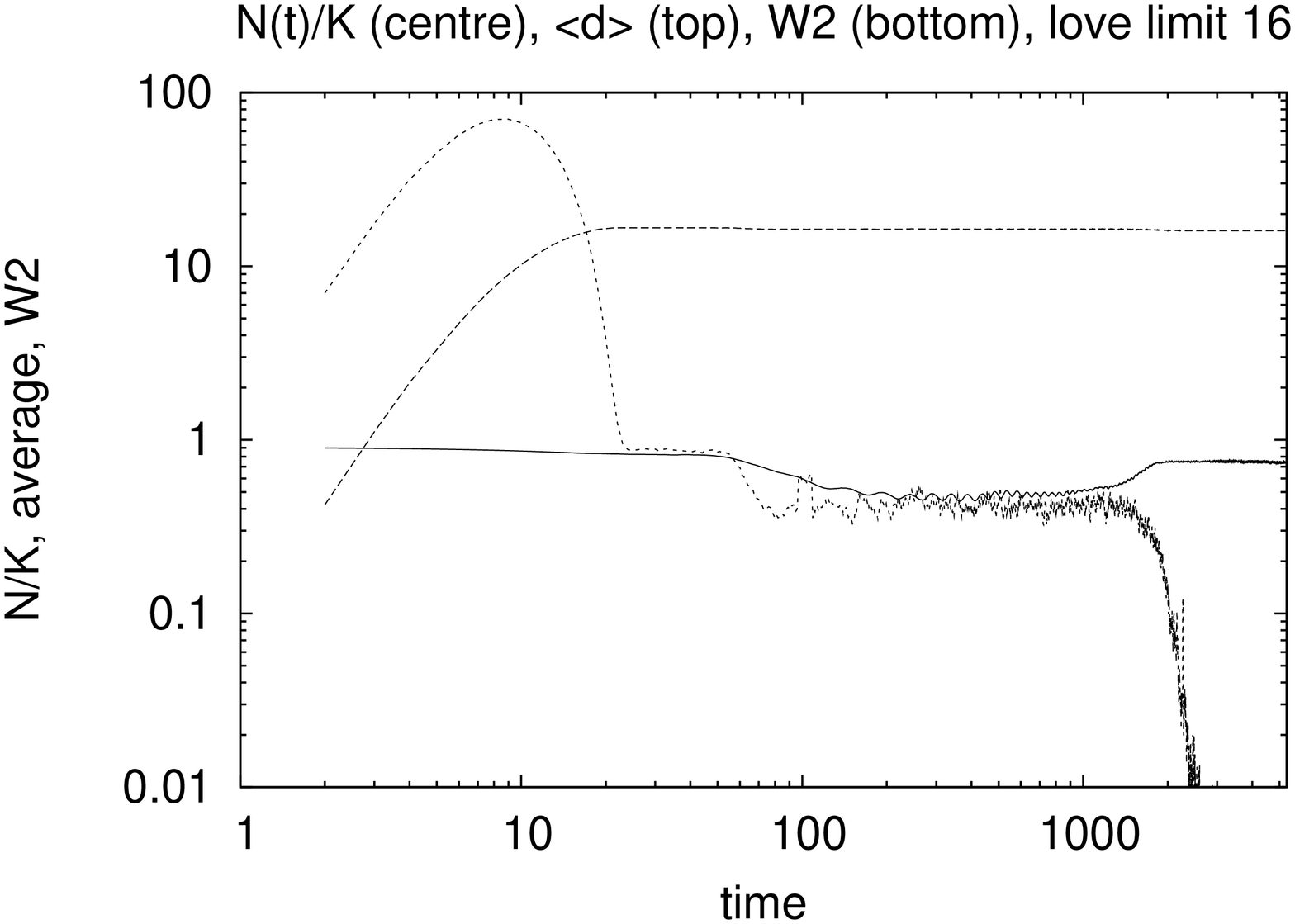}
\includegraphics[angle=-90,scale=0.4]{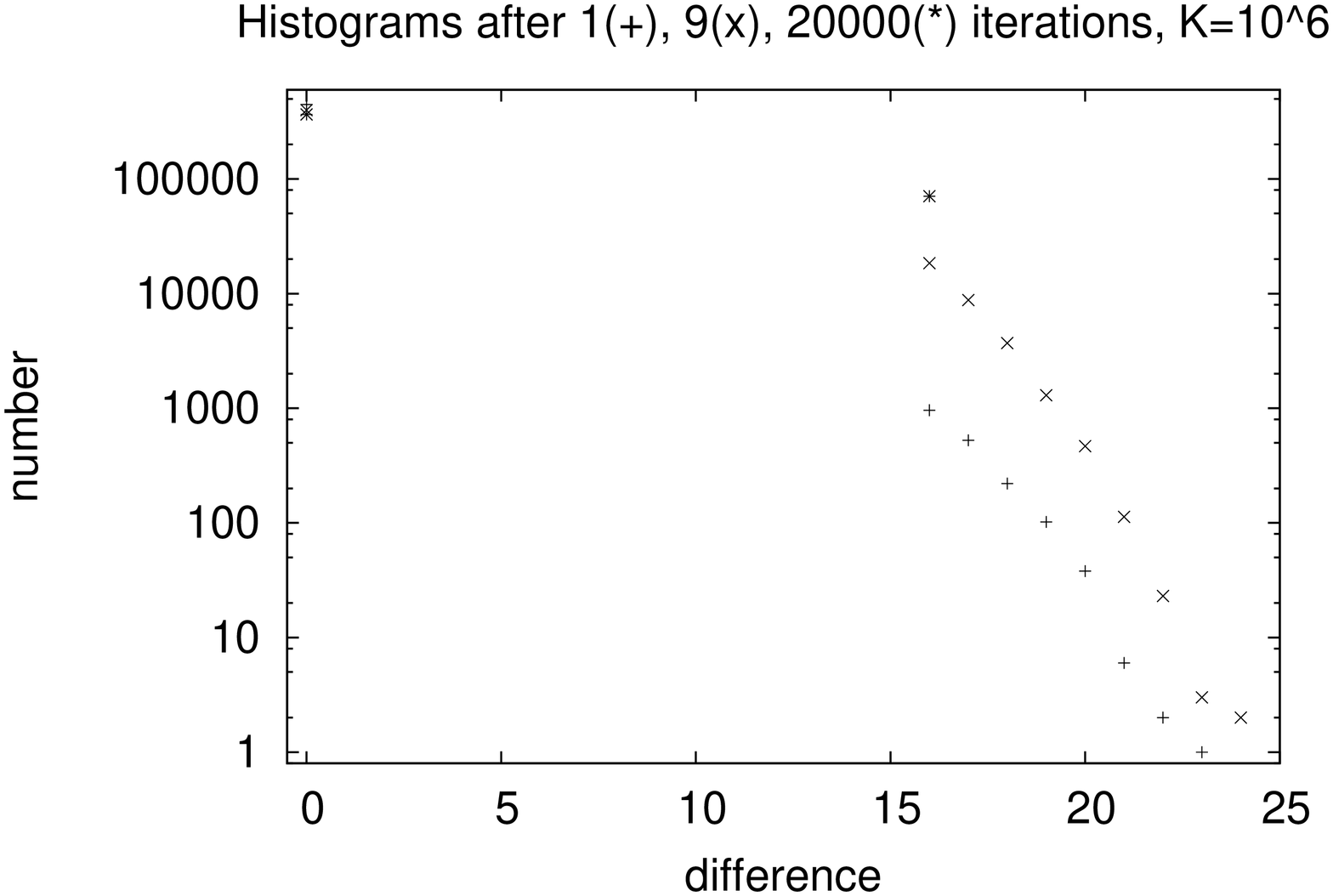}
\end{center}
\caption{
For love limit = 16, the top part 
shows the time development of the population (normalised by the carrying 
capacity $K$, central solid line), the average difference between couples (line
increasing to top), and the squared width (line going to zero after 5300 
iterations). The bottom part shows the histograms for the differences
after one, nine and 20000 iterations.
}
\end{figure}

\begin{figure}[hbt]
\begin{center}
\includegraphics[angle=-90,scale=0.4]{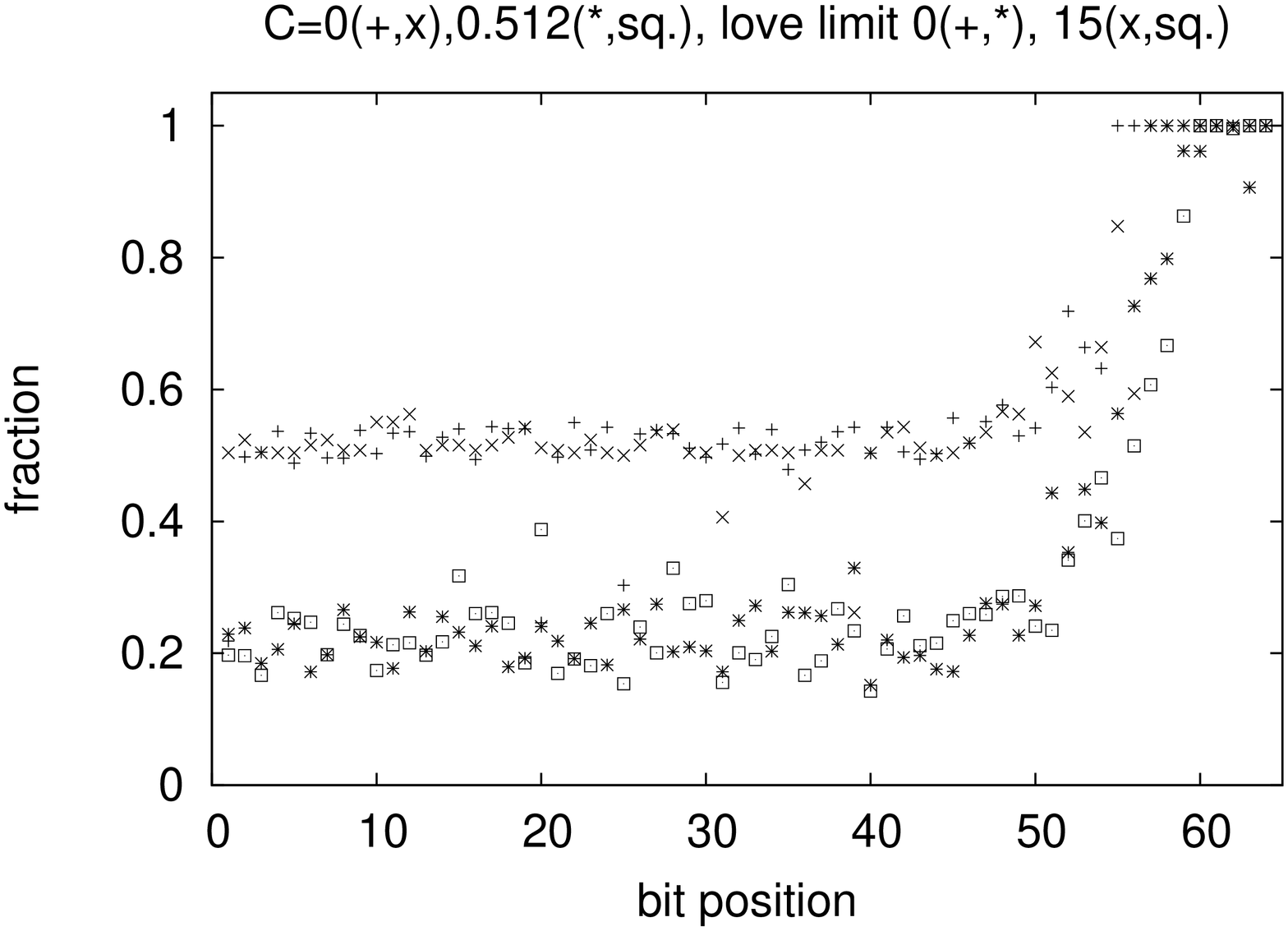}
\includegraphics[angle=-90,scale=0.4]{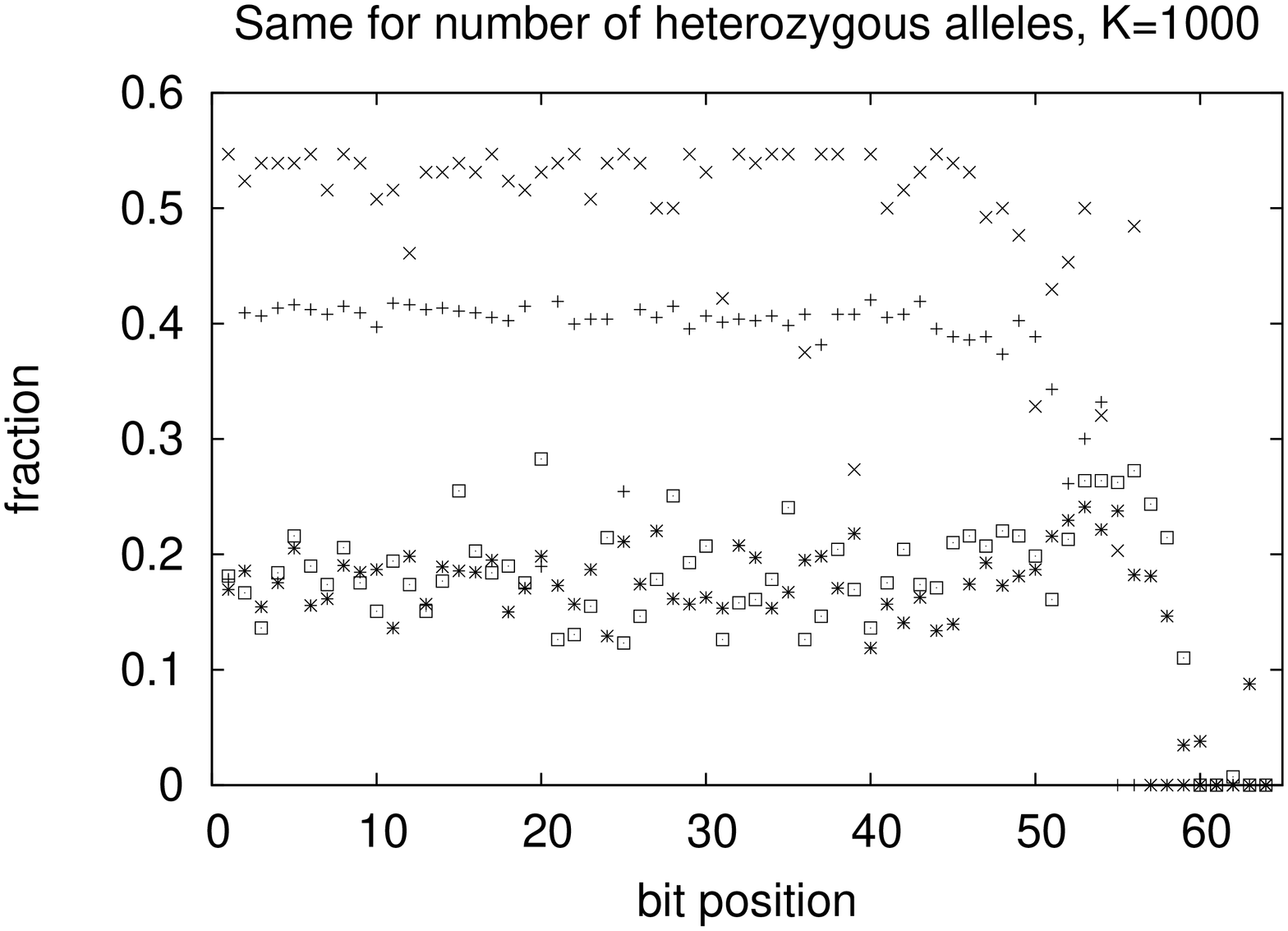}
\end{center}
\caption{
Complementarity is seen for small but not for large recombination rate $C$. 
$K=1000$, love limit 0 and 15, measured after 400 love iterations (10400 
iterations in total). The vertical scale is scaled by the population 
in the lower part and by twice the population in the upper part.
}
\end{figure}

\begin{figure}[hbt]
\begin{center}
\includegraphics[angle=-90,scale=0.4]{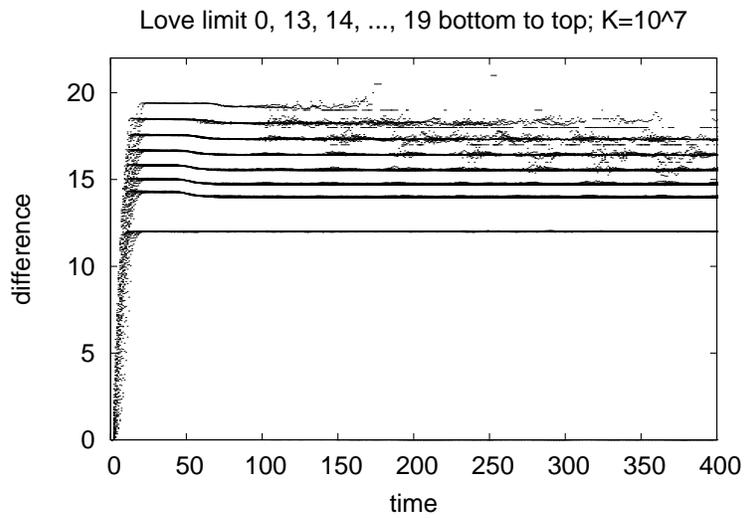}
\end{center}
\caption{
$<d>$ versus time for love limit 0, 13,14,15,16,17,18,19(extinction) bottom to 
top, and for recombination rate $C = 0, 0.001, 0.002,0.004, ... 0.512$ and 1. 
} 
\end{figure}

\begin{figure}[hbt]
\begin{center}
\includegraphics[angle=-90,scale=0.4]{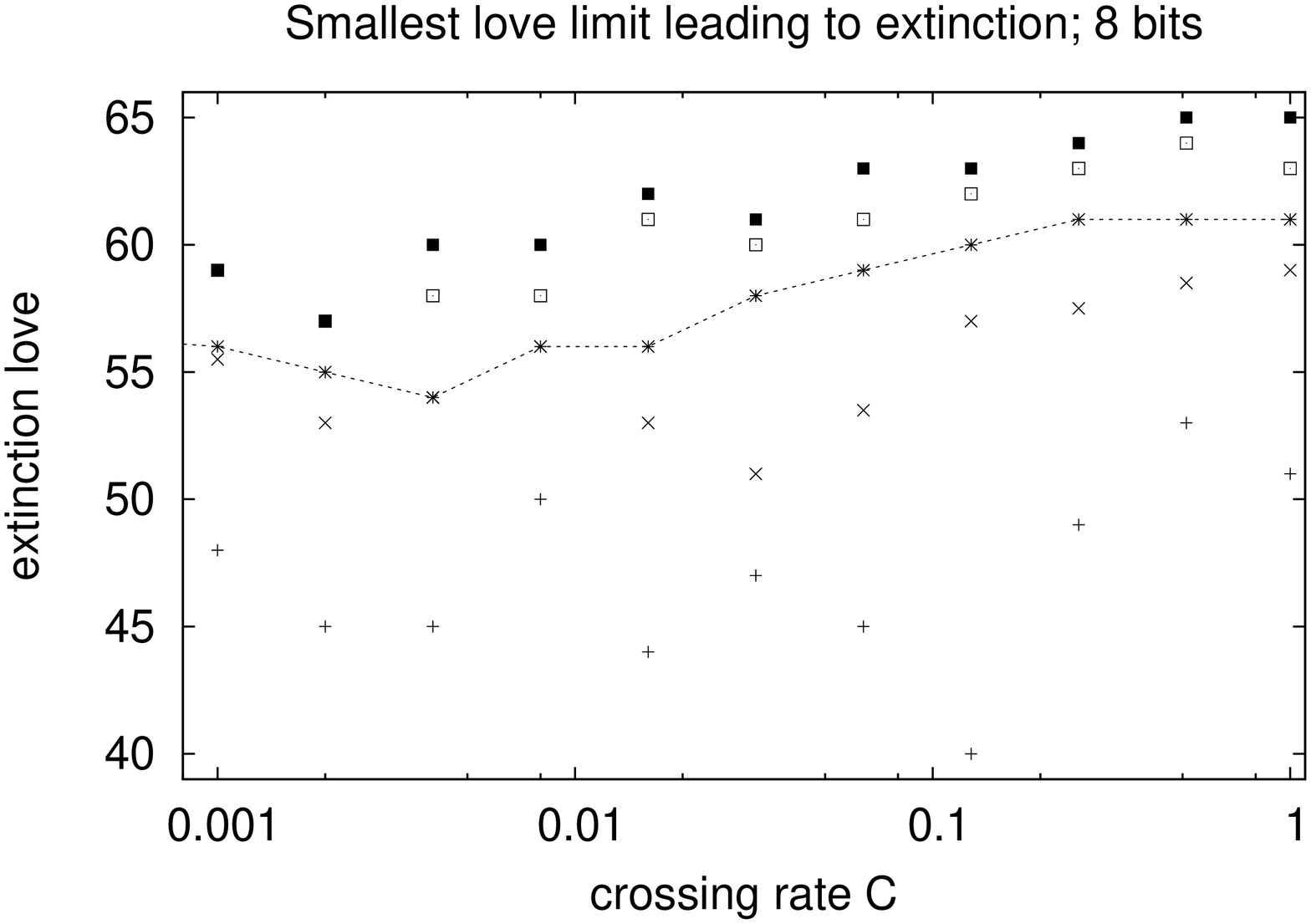}
\includegraphics[angle=-90,scale=0.4]{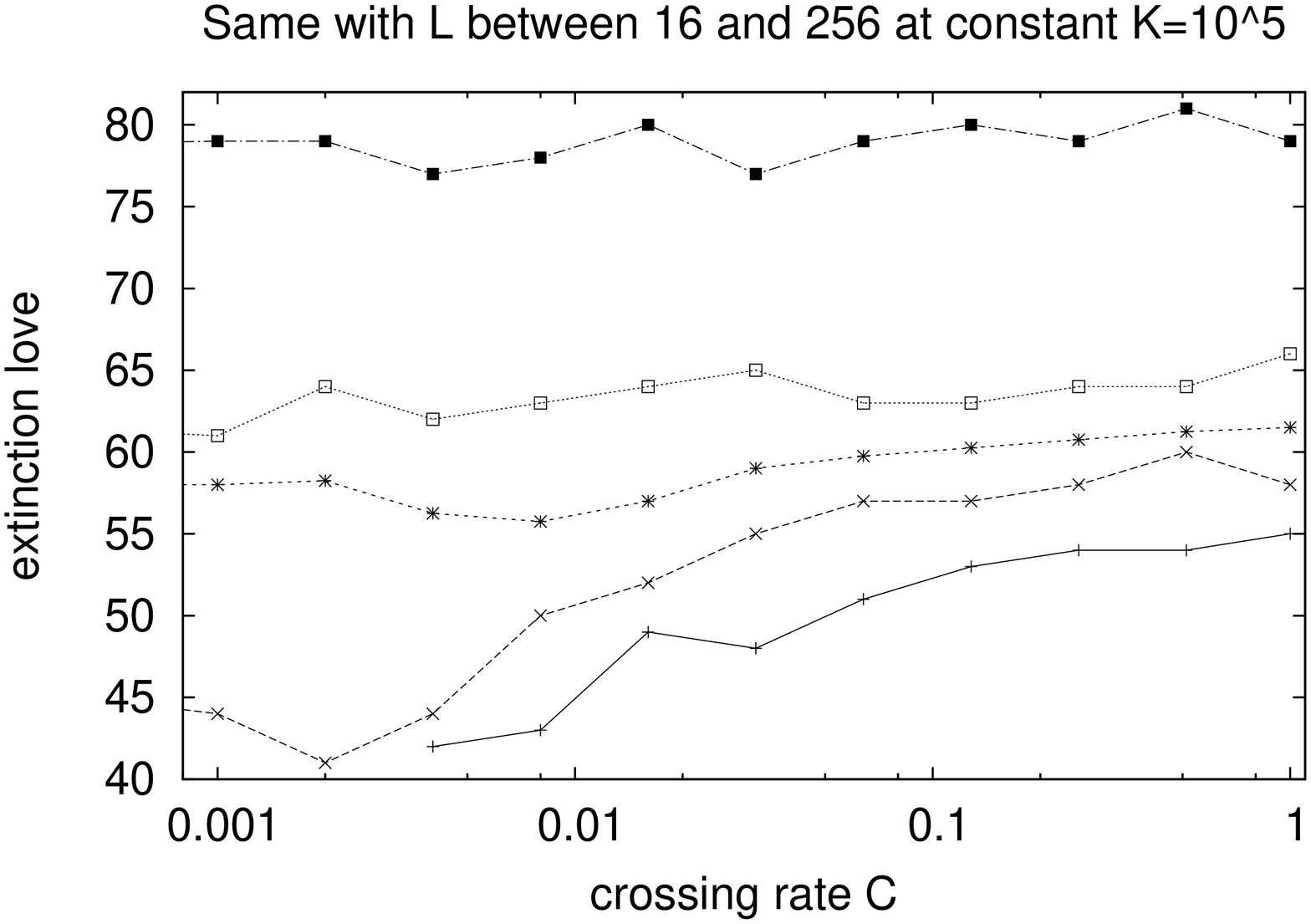}
\end{center}
\caption{ 8-bit version: 
Variation of the love limit at which first an extinction within 400 iterations
was observed, versus recombination rate $C$. Top part: $K = 10^3 \,(+), \; 10^4 
\, (\times$, average over 2 samples), $10^5$ (stars, average over 4 samples), $3
\times 10^5$ (open squares), $10^6$ (full squares). Bottom part: Genome length
$L = 16$, 32, 64, 128 and 256 from bottom to top, at $K=10^5$. 
We see that $<d>$ increases with increasing population and increasing $L$,
and is nearly independent of the recombination rate $C$. 
}
\end{figure}

\begin{figure}[hbt]
\begin{center}
\includegraphics[angle=-90,scale=0.39]{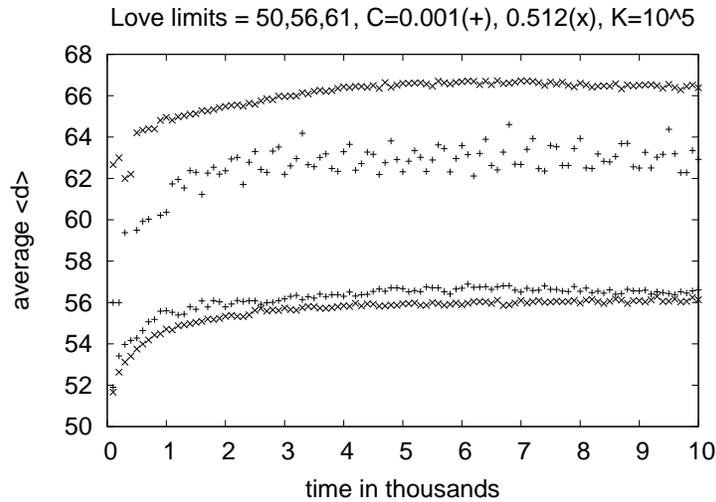}
\end{center}
\caption{8-bit version:
$<d>$ versus time for love limit and crossing rate = (50,0.512) and (50,0.001) 
(two lowest curves), (56,0.001: central curve), (60,0.512: top curve); the 
higher love limit is one unit below extinction. Not much changes for longer
times (not shown). 
}
\end{figure}

\begin{figure}[hbt]
\begin{center}
\includegraphics[angle=-90,scale=0.39]{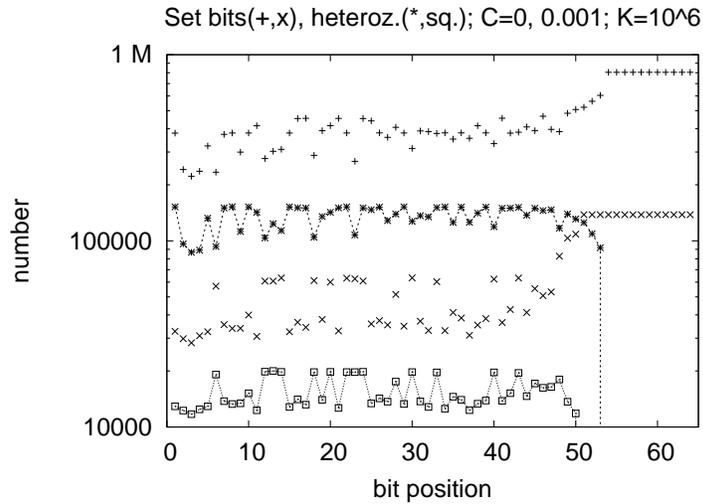}
\end{center}
\caption{8-bit version:
Complementarity versus purification for $K = 10^6$ and love limit 50: At $C = 0$
(+) about half the bits are set to one at younger ages; for $C = 
0.001$ ($\times$) this fraction is much lower. The plateaus to the 
right give twice the total population. The two lines connect the data for the 
heterozygous loci at $C=0$ ($\times$) and 0.001 (squares). 400 love iterations.
}
\end{figure}

\begin{figure}[hbt]
\begin{center}
\includegraphics[angle=-90,scale=0.39]{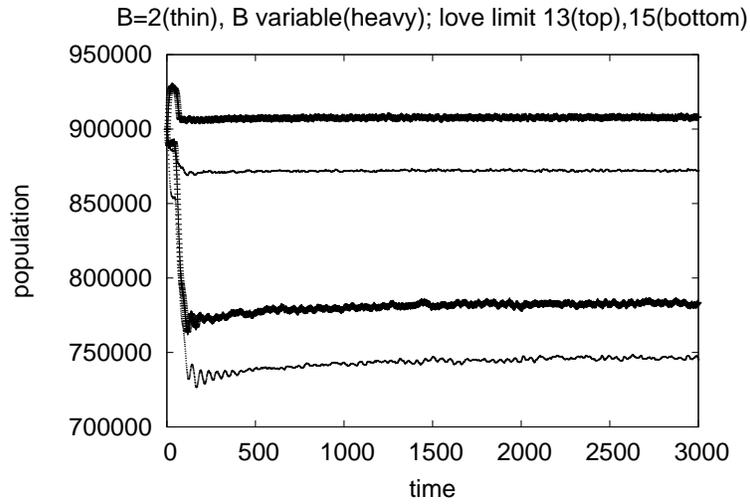}
\end{center}
\caption{One-bit version with fertility increasing with increasing
difference: $B=2+ \Delta$. The love limits = $<d> - \Delta$ (which defines the 
relative difference $\Delta$) are 13 and 15. Similar results were obtained 
in the 8-bit version using $B=2+\Delta/10$.
}
\end{figure}

\begin{figure}[hbt]
\begin{center}
\includegraphics[angle=0,scale=0.79]{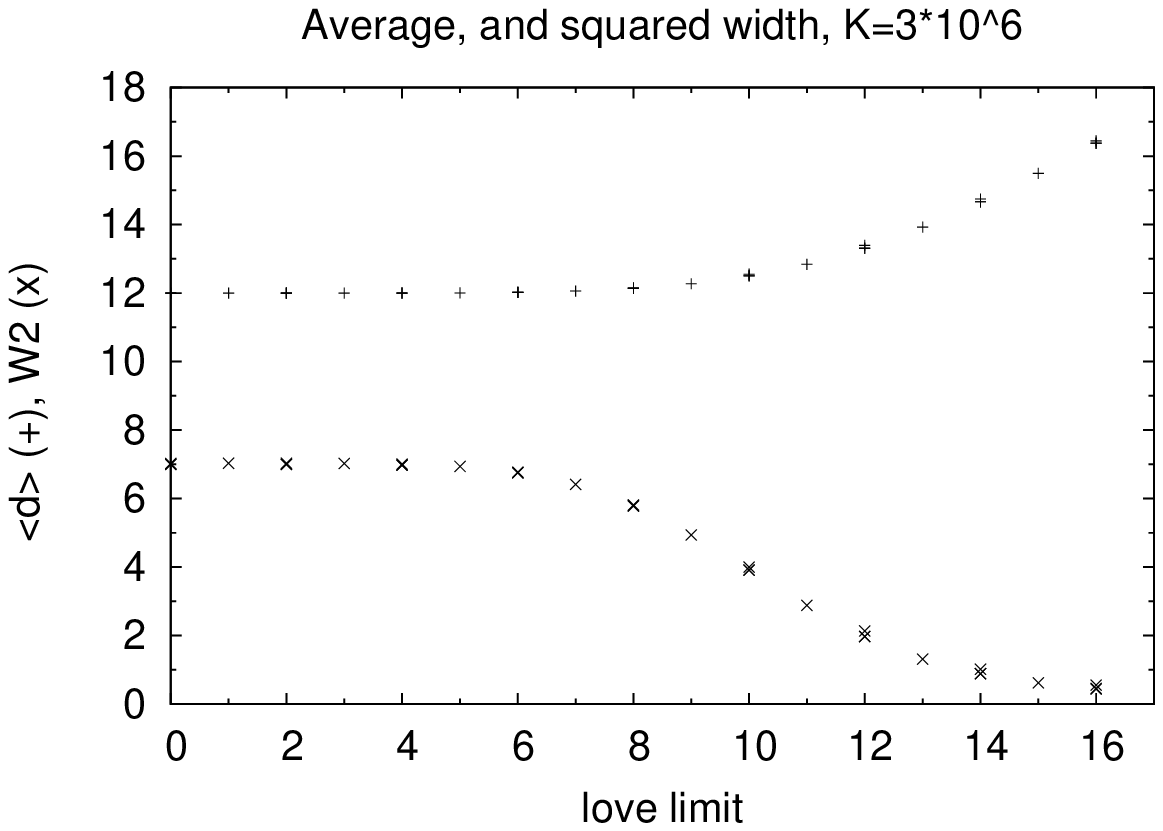}
\end{center}
\caption{Lack of influence from crossing of love strings.
}
\end{figure}

\bigskip
Fig.1 shows extinction or recovery, depending on parameters like the love limit 
which could vary between zero and 32.
The top part shows examples of long-time behaviour at $L=32$ (symbols) and
64 (line) indicating partial or full recovery after a decay within the first
few hundred time steps after love selection was switched on.
The middle part shows survival tests when our love limits for the difference 
increase from top to bottom. Extinction occurs first at a love limit of 18 
(from 32 maximal), $K= 3 \times 10^6$. (Curves for $K = 10^5, \;3 \times 10^5, 
\; 10^6$ look similar.) The bottom part shows for these and other simulations
how with increasing love limit and thus decreasing population
the average difference (+) increases while the squared width ($\times$) goes 
to zero. (Averages in the bottom part are taken from 201 to 400 iterations 
after switching on love at time 10,000.) Thus, as expected, when the love limit 
is higher (more demanding), then the average difference is higher and the 
scattering of the distances is lower. 

For longer times the differences for all couples may collapse to one value,
as seen in Fig.2 for $K = 10^6$ and love limit = 16. After 2000 iterations,
all couples have $d=16$, but most females have no partner which is counted 
as $d=0$ in the left upper corner of Fig.2b.

Thus far only the genome was mutated with one mutation per bit-string, iteration
and individual. When the same mutation rate is applied also to the love strings
then no final fixed point as in Fig.2a is reached; the average difference
always fluctuates slightly and the width of the distribution does not become
zero. (In the case of $10^7$ iterations and love limit 16 close to extinction, 
only after $10^5$ iterations was a roughly stationary equilibrium found.)
These love mutations are reversible.

If we add to these mutations the option of a recombination (or crossing) 
rate $C$ smaller than one (until now $C=1$), then complementarity \cite{zaw}
of the genome is seen for small $C$ and small populations. Fig.3 shows in its
top part the number of bits set to one as a function of bit position: Youth to 
the left, old age to the right. At the right, the maximum number (seen as a 
plateau) is twice the population since each individual has two genomic 
bit-strings. At the left we see for $C = 0$ (+ and $\times$) about half that
value as needed if at each bit position the two bits are complementary to
each other and thus without damage to the phenotype. For $C = 0.512$, in
contrast, the curves increase much steeper from left to right, meaning
that most youth bits are set to zero. This picture is confirmed by a bit-by-bit 
comparison of the genomic bit-strings within each individual (bottom part). 
Thus, love reduces the population but does not change the complementarity. 
(Here, $K = 1000$; for larger $K$ complementarity is more difficult \cite{zaw}.)
As in Fig.1b, Fig.4 shows the average difference for couples to 
increase with increasing love limit; the width again decreases (not shown);
the recombination rate $C$ has less influence on this average.

Finally, as mentioned at the beginning, instead of only two choices (one bit)
for the love alleles, we now use 8 bits (one byte) for each of the 16
elements of the two love strings (two times 128 bits in total). Thus it is
easily possible that at one locus all four alleles on the love strings 
(two in the father, two in the mother) are different. The activity
$M$ of a male allele is the number of bits set to one and varies between 0 and 
8; the same holds for the female activity $F$. A superscript (1 or 2) denotes
the two love strings. 

The difference is now

$$ d = \sum_{i=1}^{16} |M_i^1 + M_i^2 - F_i^1 - F_i^2| $$
and can vary between 0 and 256. Actually already at a love limit near $d=64$
the population dies out, Fig.5. Fig.6 shows a slow increase with time of the 
average difference; again the recombination rate has little influence on $<d>$.
Complementarity in the genome bit-strings is again observed for $C = 0$ even
at a population of 400,000, Fig.7.

Returning to our earlier love strings with 16 bits (instead of 16 bytes), 
recombination rate $C=1$, and no mutations for love strings, we now assume a
a birth rate increasing linearly with increasing difference $<d>$
of the couple, somewhat similar to \cite{berntsen}. Fig.8 shows that now the 
populations are higher than for a constant $B=2$; those for the standard Penna 
model without love are in between near 0.9 million (not shown). In this way
love can be justified by evolution if it leads to a higher effective birth rate
through paternal child care \cite{fidel} or the above-mentioned postzygotic
selection.

For this version we also introduced crossover for the love strings, in addition
to the crossover for the genomic bit-strings (one recombination per time step).
Fig.9 shows that after 400 love steps this does not matter: With or without 
crossing, with or without love mutations, the results are about the same and 
look like Fig. 1 centre. (For much longer times and love limit zero, $W2$ 
fluctuates a lot and after millions of time steps, a true equilibrium with only 
one bit-configuration in all the love strings may arise: $<d> = W2 = 0$.)

In summary, high requirements for love endanger the survival of the population.
That love nevertheless has evolved in humans proves beyond reasonable doubt
how important the father's contribution in raising the children was. 
We restricted ourselves here to love strings (corresponding to MHC and OR)
rather decoupled from genomic bit-strings; in reality both are stored in the
DNA of the chromosomes, and are linked to each other. 

We thank S. Moss de Oliveira and P.M.C. de Oliveira for many discussions of 
the dangers of love, and S.L. Kuhn for drawing our attention to Ref.1.

\end{document}